\documentclass{article}

\usepackage[utf8]{inputenc}
\usepackage[T1]{fontenc}
\usepackage{arxiv}
\usepackage{hyperref}
\usepackage{url}
\usepackage{booktabs}
\usepackage{amsfonts}
\usepackage{amsmath}
\usepackage{amssymb}
\usepackage{graphicx}
\usepackage{natbib}
\usepackage{float}
\usepackage{caption}

\title{Partitioning Israeli Municipalities into Politically Homogeneous Cantons:\\A Constrained Spatial Clustering Approach}

\author{
  Adir Elmakais \\
  Department of Computer Science\\
  Bar-Ilan University\\
  Ramat Gan, Israel\\
  \texttt{adir.elmakais@live.biu.ac.il} \\
  \And
  Oren Glickman \\
  Department of Computer Science\\
  Bar-Ilan University\\
  Ramat Gan, Israel\\
  \texttt{oren.glickman@biu.ac.il} \\
}

\date{March 2026 (revised October 2026)}

\begin{document}

\maketitle

\begin{abstract}
Israeli society has experienced significant political polarization in recent years, reflected in five Knesset elections held within a four-year period (2019--2022). Public discourse increasingly references hypothetical divisions of the country into politically homogeneous "cantons." This paper develops a data-driven algorithmic approach to explore such divisions using publicly available municipality-level election results and geographic boundary data from the Israel Central Bureau of Statistics.

We partition 255 Israeli municipalities, including the Israeli local authorities in Judea and Samaria, into cantons of neighbouring municipalities that maximize internal political similarity. Our methodology employs four clustering algorithms---Simulated Annealing, Agglomerative Clustering with contiguity constraints, Louvain Community Detection, and K-Means (baseline)---evaluated across four feature representations (BlocShares, RawParty, PCA, NMF), three distance metrics (Euclidean, Cosine, Jensen-Shannon), and six values of K (3--20), yielding 264 experimental configurations.

Key results show that BlocShares with Agglomerative clustering produces the highest clustering quality (silhouette score 0.987 with Cosine distance, 0.910 with Euclidean), while NMF with Louvain community detection achieves the best balance between political homogeneity, silhouette quality (0.142), and interpretable canton assignments. Temporal stability analysis across all five elections reveals that deterministic algorithms produce highly stable partitions (ARI 0.954--0.992 for the BlocShares configurations and 0.927 for the primary partition), while Israel's political geography remains structurally consistent despite electoral volatility. The resulting $K=5$ partition identifies five politically coherent regions---a mixed center-right metropolitan core around Tel Aviv, a right-wing region from Jerusalem through the south and the Judea and Samaria authorities, a right-leaning northern region, an Arab canton, and a canton of the seven predominantly Haredi cities, which is held together by political similarity rather than shared borders. Built from the 2019--2021 elections only, the cantons still differ significantly in every bloc's vote share in the held-out 2022 election. An interactive web application accompanies this work.
\end{abstract}


\section{Introduction}

\subsection{Motivation}

Over the past four years, Israel held five Knesset elections (April 2019, September 2019, March 2020, March 2021, and November 2022), an unprecedented frequency that reflects deep political divisions across multiple dimensions: secular versus religious, left versus right on the ideological spectrum, Arab versus Jewish communal politics, and attitudes toward judicial reform and governance. Throughout this period, public and media discourse repeatedly invoked the notion of dividing Israel into separate ``cantons'' along political lines---contrasting, for example, Tel Aviv's liberal-secular majority with Jerusalem's religious-conservative population, or the Arab-majority towns of northern Israel with predominantly Jewish communities elsewhere. The proposal was first set out in the Israeli press by \citet{strenger2014}, who argued that Israel could be divided into autonomous cantons along these lines.

While such discourse is largely rhetorical, it raises a compelling computational question: if Israel were to be divided into politically coherent regions based purely on voting patterns, what form would those regions take, and how stable would they be across election cycles? This paper applies data-driven computational methods to answer this question rigorously.

\subsection{Research Objectives}

This work has three primary objectives:

\begin{enumerate}
  \item \textbf{Partition 255 Israeli municipalities into $K$ geographically contiguous cantons} that maximize internal political homogeneity, using municipality-level election results from five Knesset elections.
  \item \textbf{Systematically compare clustering approaches} across multiple feature representations, distance metrics, algorithms, and values of $K$, evaluating trade-offs between political homogeneity, population balance, and geographic compactness.
  \item \textbf{Analyze the temporal stability} of canton boundaries across election cycles to determine whether Israel's political geography is structurally persistent or volatile.
\end{enumerate}

\subsection{Contributions}

This paper makes the following contributions:

\begin{itemize}
  \item A novel application of constrained spatial clustering to Israeli political geography, combining graph-based contiguity constraints with political feature-space optimization.
  \item A comprehensive experimental framework evaluating 264 configurations across four representations, three distance metrics, four algorithms, and six values of $K$.
  \item Temporal stability analysis across five elections using Adjusted Rand Index (ARI) and Normalized Mutual Information (NMI).
  \item Qualitative case studies validating that algorithmic cantons align with known political-demographic divisions.
  \item A held-out test of differences between cantons, using an election that did not shape the map.
  \item An interactive web application for exploring canton partitions, experiment results, and stability analysis.\footnote{\url{https://israel-cantons-project-5j7kgq8acxrbiflraqnhgb.streamlit.app/}}
  \item An open-source, modular Python codebase enabling reproducible analysis and extension.\footnote{\url{https://github.com/adirelm/israel-cantons-project}}
\end{itemize}

\subsection{Paper Outline}

Section~\ref{sec:related} reviews related work. Section~\ref{sec:problem} formally defines the constrained canton partitioning problem. Section~\ref{sec:data} describes the data sources and processing pipeline. Section~\ref{sec:method} presents the methodology. Section~\ref{sec:experiments} reports experimental results. Section~\ref{sec:stability} analyzes temporal stability. Section~\ref{sec:casestudies} presents qualitative case studies. Section~\ref{sec:limitations} discusses limitations and future work, Section~\ref{sec:ethics} addresses ethical considerations, and Section~\ref{sec:conclusion} concludes.

\paragraph{Revision (October 2026).} This version corrects the party-to-bloc mapping (the Religious Zionist Party in 2021, Otzma Yehudit in 2020 and Jewish Home in 2022 had not been assigned to a bloc); adds the 23 Israeli local authorities in Judea and Samaria and three municipalities inside the Green Line that had been lost to name mismatches; computes silhouette scores in each configuration's own distance metric; reports voter-weighted canton profiles and the primary configuration's own cross-election stability; makes Simulated Annealing reproducible for a given seed; adds per-election maps and a held-out test; and corrects the comparison with administrative districts. With the added municipalities the seven predominantly Haredi cities form a canton of their own and the two Arab cantons of the first version merge.

\section{Related Work}
\label{sec:related}

\subsection{Electoral Redistricting and Gerrymandering}

The problem of dividing a territory into politically meaningful regions has deep connections to electoral redistricting research. In democratic systems with district-based representation, electoral boundaries must be periodically redrawn to reflect population changes, and the drawing process can be manipulated to favor particular parties---a practice known as gerrymandering~\citep{stephanopoulos2015}.

\citet{brieden2017} proposed a constrained clustering approach to electoral redistricting based on generalized Voronoi diagrams and linear programming. Their method partitions geographic units into districts with hard population balance constraints while minimizing intra-district political heterogeneity.

Our work differs from classical redistricting in important ways: (1) we aim to \emph{maximize} political homogeneity within cantons, whereas fair redistricting often seeks competitive or balanced districts; (2) we operate at the municipality level rather than individual census blocks; and (3) our work is purely descriptive and exploratory rather than prescriptive. However, the shared technical challenges---ensuring geographic contiguity, handling population constraints, and operating on spatial graphs---make redistricting research directly relevant.

Recent computational approaches to fair redistricting include Markov Chain Monte Carlo methods that sample from the space of valid partitions to detect gerrymandering outliers~\citep{deford2021}, and graph-based metrics for measuring district compactness~\citep{duchin2018}.

\subsection{Spatial and Constrained Clustering}

Standard clustering algorithms---K-Means, hierarchical agglomerative methods, spectral clustering---operate in feature space without regard to geographic structure~\citep{hastie2009}. When geographic contiguity is required (as in regionalization), these methods must be extended with spatial constraints.

Spatially-constrained agglomerative clustering is a natural approach: starting with each spatial unit as its own cluster, iteratively merge adjacent clusters that are most similar in feature space~\citep{murtagh2014}. Spectral clustering leverages the graph Laplacian to embed graph nodes into a low-dimensional space before applying K-Means~\citep{luxburg2007}.

Community detection algorithms from network science, particularly the Louvain method~\citep{blondel2008}, offer an alternative by maximizing modularity~\citep{newman2004}. Metaheuristic approaches such as Simulated Annealing can optimize arbitrary multi-objective cost functions over the partition space while maintaining contiguity through constrained neighborhood operations~\citep{kirkpatrick1983}.

\subsection{Political Geography of Israel}

Israel's political landscape is characterized by several cross-cutting cleavages~\citep{diskin2005}: the left-right ideological spectrum, the secular-religious divide, the Arab-Jewish divide, and strong spatial autocorrelation---nearby municipalities tend to vote similarly due to shared demographics and social networks~\citep{arian2008, shamir1999, anselin1995}. This geographic structure makes Israel a suitable testbed for constrained spatial clustering.

\subsection{Dimensionality Reduction for Voting Data}

Municipality-level election data can be represented as high-dimensional vectors of party vote shares. Dimensionality reduction techniques compress these vectors while preserving key political structure: PCA finds orthogonal axes of maximum variance~\citep{jolliffe2002}, NMF decomposes vote shares into additive, interpretable ``political archetypes''~\citep{lee1999}, and manual bloc aggregation groups parties into predefined political blocs, producing a low-dimensional but domain-informed representation.

\subsection{Research Gap}

Prior work on constrained spatial clustering has focused primarily on fair electoral redistricting---producing \emph{competitive} districts with balanced partisan composition~\citep{stephanopoulos2015, brieden2017, deford2021}. In contrast, our work seeks to maximize political \emph{homogeneity} within regions. While regionalization methods such as SKATER~\citep{assuncao2006} and MAX-P-REGIONS~\citep{duque2012} address spatially constrained partitioning, they have not been applied to Israeli political geography, nor evaluated across a systematic grid of feature representations, distance metrics, and algorithms. Existing studies of Israeli political geography~\citep{arian2008, shamir1999} are primarily qualitative. This work bridges these gaps by combining constrained spatial clustering with a comprehensive experimental comparison tailored to the Israeli political context.

\section{Problem Definition}
\label{sec:problem}

\subsection{Formal Statement}

Let $V = \{v_1, \ldots, v_n\}$ be a set of $n$ municipalities, each characterized by a political feature vector $f_i \in \mathbb{R}^d$ derived from election results. Let $G = (V, E)$ be a geographic contiguity graph where edge $(v_i, v_j)$ exists if municipalities $i$ and $j$ share a geographic boundary. Let $w_i$ denote the voter population of municipality $i$.

\textbf{The Canton Partitioning Problem:} Given $V$, $G$, feature vectors $\{f_i\}$, voter weights $\{w_i\}$, and a target number of cantons $K$, find a partition $\mathcal{C} = \{C_1, \ldots, C_K\}$ of $V$ that:

\begin{enumerate}
  \item \textbf{Maximizes political homogeneity:} Minimizes within-canton political variance.
  \item \textbf{Ensures contiguity:} Each canton $C_k$ induces a connected subgraph of the analysis graph $G$ (shared borders plus the links added during graph preprocessing).
  \item \textbf{Balances population:} Minimizes disparity in total voter population across cantons.
\end{enumerate}

This is a variant of the regionalization problem~\citep{assuncao2006, duque2012}.

\subsection{Multi-Objective Formulation}

These objectives may conflict. We adopt a weighted multi-objective approach:

\begin{equation}
  \text{Cost}(\mathcal{C}) = \alpha \cdot \text{Homogeneity}(\mathcal{C}) + \beta \cdot \text{Balance}(\mathcal{C}) + \gamma \cdot \text{Compactness}(\mathcal{C})
  \label{eq:cost}
\end{equation}

where $\alpha, \beta, \gamma$ are tunable weights. The three components are defined as:

\begin{itemize}
  \item $\text{Homogeneity}(\mathcal{C}) = \sum_{k=1}^{K} \frac{|C_k|}{n} \cdot \overline{\text{Var}}(\mathbf{f}_i : i \in C_k)$, the population-weighted average of within-canton feature variance across dimensions.
  \item $\text{Balance}(\mathcal{C}) = \sigma(\mathbf{w}) / \mu(\mathbf{w})$, the coefficient of variation of canton voter populations $\mathbf{w} = (w_1, \ldots, w_K)$.
  \item $\text{Compactness}(\mathcal{C}) = |\{(u,v) \in E : C(u) \neq C(v)\}| / |E_{\text{assigned}}|$, the fraction of graph edges crossing canton boundaries (cut ratio), where $E_{\text{assigned}}$ is the set of edges whose both endpoints have been assigned to a canton.
\end{itemize}

In our implementation, we use $\alpha = 0.4$, $\beta = 0.4$, $\gamma = 0.2$, reflecting equal priority for homogeneity and population balance with compactness as a secondary regularizer. Feature vectors are standardized (zero-mean, unit-variance) before cost computation.

\subsection{Contiguity Constraint}

The contiguity constraint requires that each canton $C_k$ forms a connected subgraph of the analysis graph $G$; because $G$ also links isolated towns and same-bloc enclaves, a canton can include towns that do not border it. This is enforced structurally in agglomerative methods (only adjacent clusters merge) and algorithmically in SA (moves are rejected if they break contiguity).

\section{Data Sources}
\label{sec:data}

\subsection{Election Data}

We use municipality-level election results from five consecutive Knesset elections, obtained from the Israel Central Bureau of Statistics (CBS)~\citep{cbs2023elections}. Table~\ref{tab:elections} summarizes the data scope.

\begin{table}[htbp]
  \centering
  \caption{Municipality-level election data across five Knesset elections (all cover 255 municipalities).}
  \label{tab:elections}
  \begin{tabular}{clcccc}
    \toprule
    Election & Knesset & Date & Eligible & Actual & Turnout \\
    \midrule
    1 & 21 & Apr.\ 2019 & 6,303,973 & 4,089,609 & 64.9\% \\
    2 & 22 & Sep.\ 2019 & 6,354,847 & 4,166,925 & 65.6\% \\
    3 & 23 & Mar.\ 2020 & 6,416,082 & 4,268,189 & 66.5\% \\
    4 & 24 & Mar.\ 2021 & 6,539,675 & 3,998,159 & 61.1\% \\
    5 & 25 & Nov.\ 2022 & 6,742,445 & 4,311,984 & 64.0\% \\
    \bottomrule
  \end{tabular}
\end{table}

\subsection{Geographic Data}

Municipal boundary polygons were obtained from the Israel CBS geographic databases~\citep{cbs2023geo} in Shapefile format. Municipalities were dissolved from locality-level polygons to municipality-level using an official locality-to-municipality mapping. Three municipalities carried outdated names in the boundary file and were missing from the first version of this work: Nof HaGalil (still named Nazareth Illit), the Sdot Dan regional council (named Emek Lod) and Shibli-Umm al-Ghanam (spelled differently); all three are now matched.

\textbf{Judea and Samaria.} Election results include the 23 Israeli local authorities in Judea and Samaria, but the CBS file has no polygons for them. The 17 cities and local councils (including Ariel, Ma'ale Adumim, Beitar Illit and Modi'in Illit) come from the CBS municipal layer based on the Ministry of Interior's boundaries as of the end of 2023~\citep{cbs2023geo}. The 6 regional councils (Shomron, Mateh Binyamin, Gush Etzion, Har Hevron, Megilot and Arvot HaYarden) come from the Civil Administration's council-area and settlement-jurisdiction layers, published under freedom of information~\citep{civiladmin2026}: each council is the union of its own area and its member settlements' jurisdictions, minus the cities and local councils. Where council areas overlap in the source (79~km$^2$ between Mateh Binyamin and Arvot HaYarden), the overlap goes to the council with more of its settlements' area inside it, or, with none on either side, to the council for which it is the larger share of its area. Invalid source geometries are repaired, new polygons are clipped so they do not overlap existing ones, and all new polygons are simplified together by 5~m so shared borders stay shared. Points inside Ramallah, Al-Bireh, Nablus, Jericho, Bethlehem and Hebron's old city fall outside every polygon: the analysis covers Israeli local authorities, whose jurisdictions in Judea and Samaria are fragmented, while Palestinian areas, which do not vote in Knesset elections, are outside it.

The combined file has 257 municipalities, of which 255 have votes in all five elections and form the analysis set; the other two are industrial local councils with no voters.

\subsection{Party-to-Bloc Mapping}

Israeli political parties were classified into five political blocs, as shown in Table~\ref{tab:blocs}. Every list with at least 1\% of the valid vote in an election is assigned to a bloc; smaller lists count as ``other''. The mapping is keyed by ballot letter per election, since letters move between parties (in 2021 Yamina ran under the letter Bet and the Religious Zionist Party under Tet).

\begin{table}[htbp]
  \centering
  \caption{Party-to-bloc classification.}
  \label{tab:blocs}
  \begin{tabular}{llp{6.2cm}}
    \toprule
    Bloc & Description & Example Parties \\
    \midrule
    Right & Nationalist right & Likud, Yisrael Beiteinu, Yamina, Religious Zionism, Otzma Yehudit, Jewish Home \\
    Haredi & Ultra-Orthodox religious & Shas, United Torah Judaism \\
    Center & Centrist parties & Yesh Atid, Blue and White \\
    Left & Labor-left & Labor, Meretz \\
    Arab & Arab-majority parties & Joint List, Ra'am, Balad \\
    \bottomrule
  \end{tabular}
\end{table}

\subsection{Data Processing Pipeline}

\begin{figure}[htbp]
  \centering
  \includegraphics[width=0.95\linewidth]{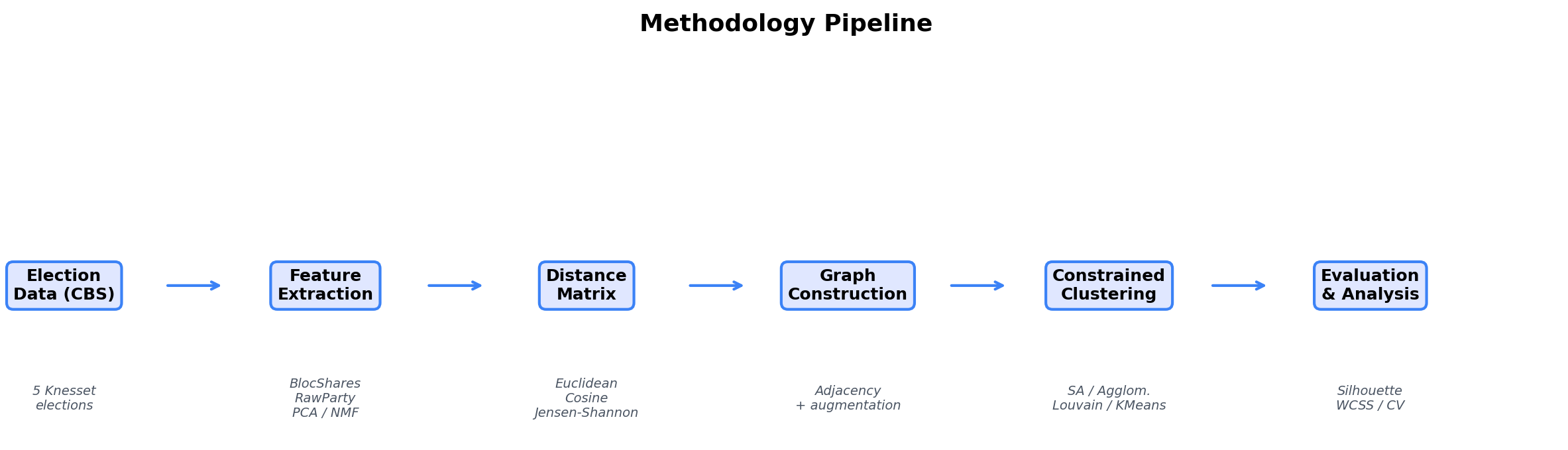}
  \caption{End-to-end methodology pipeline from raw election data to evaluated canton partitions.}
  \label{fig:pipeline}
\end{figure}

The data processing pipeline (Figure~\ref{fig:pipeline}) handles: (1)~loading raw election files and geographic shapefiles; (2)~name normalization for cross-dataset matching; (3)~municipality dissolution from locality-level to municipality-level polygons; and (4)~feature computation from raw vote counts to vote share vectors.

\section{Methodology}
\label{sec:method}

\subsection{Feature Representations}

We evaluate four feature representations for municipality political profiles:

\textbf{BlocShares (11 features):} For each municipality, compute the mean and standard deviation of each bloc's vote share across all five elections, plus the average voter count. This produces an 11-dimensional vector capturing both the political profile and its temporal stability.

\textbf{RawParty (51 features):} The vote share of every list that won seats (passed the 3.25\% threshold) in every election, one feature per list and election (11, 9, 8, 13 and 10 lists in the five elections). Elections are kept apart rather than averaged, because a ballot letter can stand for different parties in different elections, and no party-to-bloc mapping is used; with every list included (186 features), 8 of the 255 municipalities change canton in the primary map. The canton partition therefore rests on how municipalities voted, not on how parties are grouped; blocs describe the cantons afterwards and enter the method only through the enclave edges of the graph.

\textbf{PCA\_5 (5 features):} Five principal components extracted from the RawParty representation via PCA~\citep{jolliffe2002}, capturing the dominant axes of political variation.

\textbf{NMF\_5 (5 features):} Five components from Non-negative Matrix Factorization~\citep{lee1999} of the RawParty representation, producing additive, non-negative components interpretable as political archetypes.

\subsection{Distance Metrics}

Three distance metrics measure political dissimilarity between municipalities:

\textbf{Euclidean Distance:} The standard $L_2$ norm between feature vectors. Sensitive to magnitude differences.

\textbf{Cosine Distance:} Measures angular distance between feature vectors ($1 - \cos\theta$). Invariant to vector magnitude, focusing on relative proportions rather than scale.

\textbf{Jensen-Shannon Distance:} The square root of the Jensen-Shannon divergence~\citep{lin1991}, a symmetric, bounded measure for probability distributions:

\begin{equation}
  \text{JSD}(P \| Q) = \frac{1}{2} D_{KL}(P \| M) + \frac{1}{2} D_{KL}(Q \| M), \quad M = \frac{1}{2}(P + Q)
  \label{eq:jsd}
\end{equation}

Our implementation uses $d(P, Q) = \sqrt{\text{JSD}(P \| Q)}$ (Eq.~\ref{eq:jsd}), which is a proper metric~\citep{endres2003}.

\subsection{Graph Construction and Preprocessing}

We construct a contiguity graph $G = (V, E)$ where municipalities are nodes and edges connect municipalities sharing a geographic boundary. The raw graph has 257 nodes and 574 edges but is \emph{disconnected}. To ensure a connected graph, we apply three preprocessing steps:

\begin{enumerate}
  \item \textbf{Virtual edges for isolates:} For each isolated node, connect it to its $k=3$ nearest neighbors in feature space.
  \item \textbf{Enclave edges:} For municipalities where a single bloc exceeds 70\% of votes, add edges connecting all enclaves of the same bloc type.
  \item \textbf{Bridge edges:} For remaining disconnected components, connect each to the largest component via the most politically similar pair.
\end{enumerate}

Municipalities inside the Green Line are adjacent when their polygons share a boundary line. Polygons from different sources do not share exact borders, so a Judea and Samaria authority is adjacent to any polygon that runs alongside it within 100~m (their 100~m buffers overlap by at least $200 \times 200$~m$^2$); Ariel and Elkana, which touch nothing within that distance, are joined to their geographically nearest authority (Shomron). After subsetting to the 255 analysis municipalities (569 edges), augmentation produces a fully connected graph with 255 nodes and 2,490 edges.

\subsection{Clustering Algorithms}

\textbf{Simulated Annealing (SA):} SA optimizes the multi-objective cost function (Eq.~\ref{eq:cost}) over the partition space. Starting from a seed-based initialization (each canton is seeded with the municipality having the highest vote share for the corresponding political bloc), the algorithm proposes moves by reassigning a border municipality to an adjacent canton. Moves that improve the cost are always accepted; moves that increase cost by $\Delta$ are accepted with probability $\exp(-\Delta/T)$ (Metropolis criterion~\citep{kirkpatrick1983}), where $T$ is the current temperature. Moves that would break contiguity are rejected. Parameters: initial temperature $T_0 = 1.0$, cooling rate $0.9995$, 5,000 iterations per configuration (a grid-search budget constraint; see Section~\ref{sec:limitations} for a 50,000-iteration sensitivity analysis).

\textbf{Agglomerative Clustering:} Average-linkage with contiguity constraints~\citep{murtagh2014}---starting from 255 singleton clusters, iteratively merge the two adjacent clusters with the smallest average pairwise distance. Deterministic and contiguity-preserving by construction.

\textbf{Louvain Community Detection:} Maximizes modularity~\citep{newman2004, blondel2008} on the adjacency graph with edge weights derived from political similarity: $w(u,v) = 1 - d(u,v)/d_{\max}$, where $d$ is the chosen distance metric. Louvain does not directly accept a target $K$; our implementation performs a binary search over the resolution parameter $r \in [0.01, 10.0]$ (up to 30 iterations, tolerance $10^{-6}$) to find the resolution that produces the number of communities closest to the desired $K$.

\textbf{K-Means Baseline:} Standard K-Means~\citep{hastie2009} in feature space without geographic constraints. Cantons may be geographically disconnected.

\subsection{Evaluation Metrics}

\textbf{Silhouette Score}~\citep{rousseeuw1987}: Measures how well each municipality fits in its assigned canton versus the nearest alternative. Ranges from $-1$ to $+1$.

\textbf{Within-Canton Sum of Squares (WCSS):} Total political variance within cantons, $\text{WCSS} = \sum_{k} \sum_{i \in C_k} \|\mathbf{f}_i - \boldsymbol{\mu}_k\|^2$. WCSS trends closely follow silhouette scores; we report silhouette as the primary quality metric for conciseness.

\textbf{Population Balance (CV):} Coefficient of variation of canton voter populations.

\textbf{Contiguity:} Number of cantons that are disconnected in the adjacency graph. All graph-based algorithms (SA, Agglomerative, Louvain) enforce contiguity by construction. K-Means, operating without geographic constraints, produces disconnected cantons in the majority of configurations.

All algorithms were implemented in Python using scikit-learn~\citep{pedregosa2011}, NetworkX~\citep{hagberg2008}, and SciPy~\citep{virtanen2020}.

\section{Experimental Results}
\label{sec:experiments}

\subsection{Experimental Setup}

We conducted a systematic grid search over 264 configurations (Table~\ref{tab:grid}):

\begin{table}[htbp]
  \centering
  \caption{Experimental grid dimensions.}
  \label{tab:grid}
  \begin{tabular}{llc}
    \toprule
    Dimension & Values & Count \\
    \midrule
    Representation & BlocShares, RawParty, PCA\_5, NMF\_5 & 4 \\
    Distance Metric & Euclidean, Cosine, Jensen-Shannon & 3 \\
    Algorithm & SA, Agglomerative, Louvain, K-Means & 4 \\
    $K$ (cantons) & 3, 5, 7, 10, 15, 20 & 6 \\
    \textbf{Total} & $4 \times 3 \times 4 \times 6 = 288$ minus 24 excluded$^*$ & \textbf{264} \\
    \bottomrule
  \end{tabular}

  \smallskip
  {\footnotesize $^*$Jensen-Shannon divergence requires non-negative inputs and is incompatible with PCA\_5 (which produces negative values); all 24 PCA\_5/JSD combinations are excluded.}
\end{table}

All 264 configurations executed successfully with 0 failures.

\subsection{Key Findings}

\begin{figure}[htbp]
  \centering
  \includegraphics[width=0.8\linewidth]{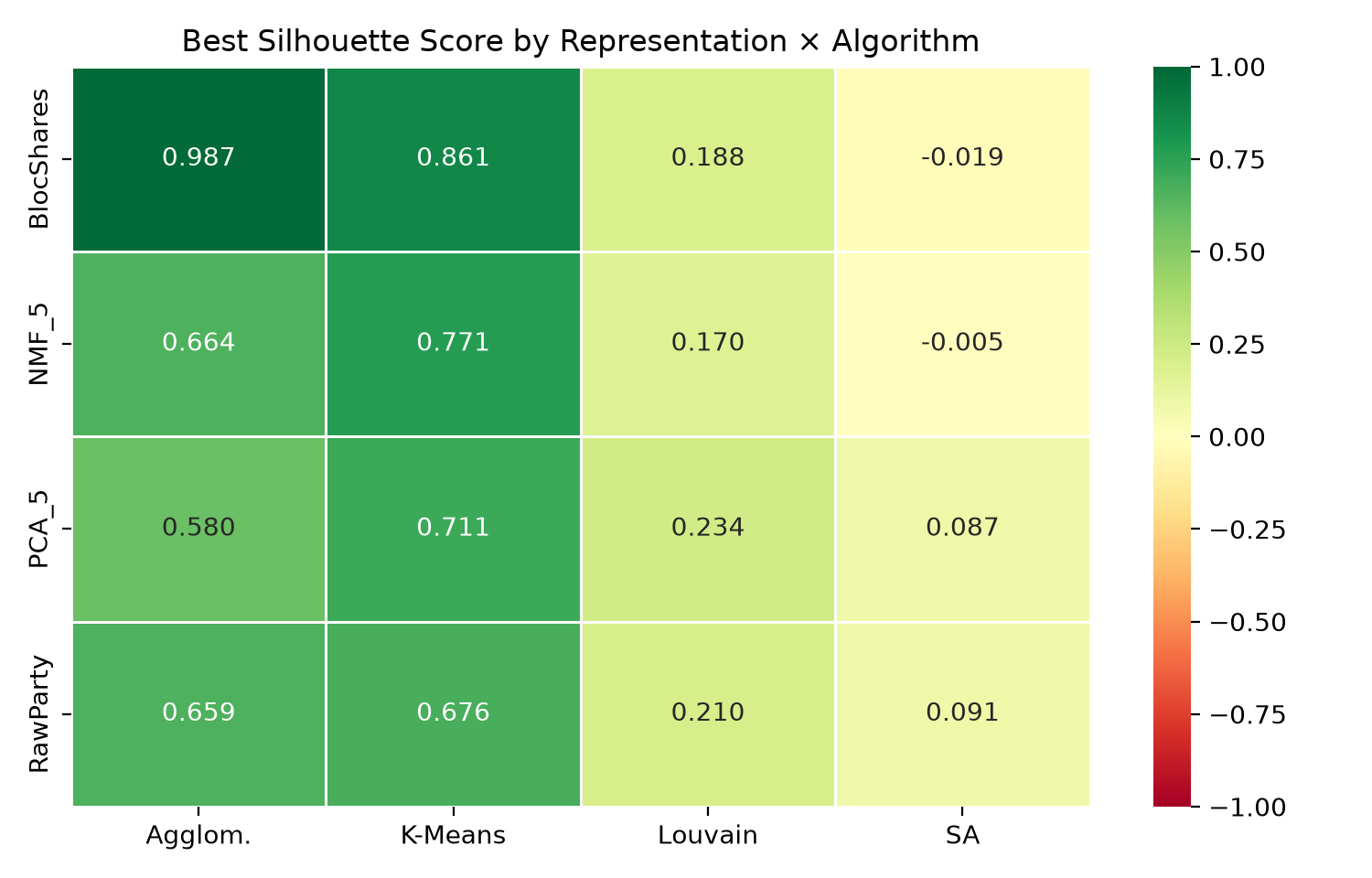}
  \caption{Best silhouette score achieved by each representation $\times$ algorithm combination.}
  \label{fig:heatmap}
\end{figure}

\textbf{Representation Comparison:} BlocShares achieves the highest silhouette scores for unconstrained algorithms (Figure~\ref{fig:heatmap}), thanks to its domain-informed aggregation. However, with contiguity-constrained SA, BlocShares produces severely imbalanced partitions. The party-level representations (RawParty and the NMF\_5 and PCA\_5 components derived from it) capture finer-grained political distinctions~\citep{lee1999} and give better-balanced SA partitions. RawParty and NMF\_5 perform similarly (best silhouettes 0.210 and 0.170 with Louvain); NMF\_5 is used for the primary map because its five non-negative components are easier to interpret, and low-dimensional features suit distance-based clustering~\citep{bellman1961}.

\textbf{Distance Metric Comparison:} Silhouette scores are computed in each configuration's own distance metric. Cosine distance produces the highest silhouette overall (0.987 with BlocShares/Agglomerative at $K=3$), ahead of Euclidean (0.910) and Jensen-Shannon (0.793) for the same configuration. Metric choice matters most for the K-Means baseline, which optimizes Euclidean distance regardless of the metric used for evaluation: its $K=3$ BlocShares partition scores 0.861 under Euclidean distance but $-$0.864 under Cosine and $-$0.199 under Jensen-Shannon.

\textbf{Algorithm Comparison:}
\begin{itemize}
  \item \textbf{Agglomerative} achieves the highest silhouette scores (up to 0.987) but can produce severe population imbalance.
  \item \textbf{SA} produces the most balanced cantons due to its explicit population balance objective, at the cost of lower silhouette scores.
  \item \textbf{Louvain} reaches the target $K$ by searching over its resolution parameter (exactly in 60 of 66 runs, off by one in the rest); highly stable across elections (ARI 0.927--0.992).
  \item \textbf{K-Means} achieves competitive silhouette scores but produces geographically disconnected cantons.
\end{itemize}

\begin{figure}[htbp]
  \centering
  \includegraphics[width=0.8\linewidth]{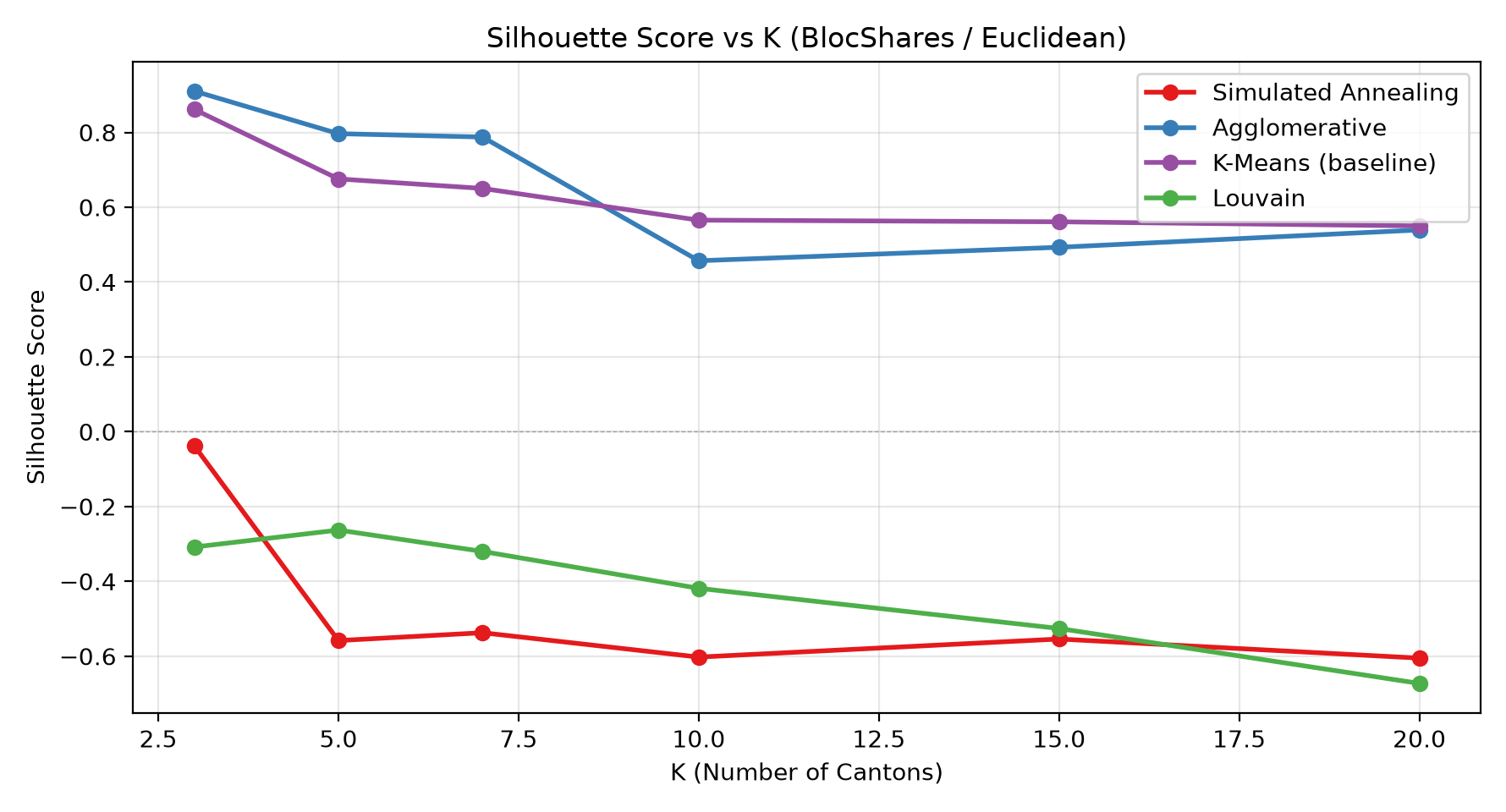}
  \caption{Silhouette score vs $K$ for BlocShares/Euclidean across all four algorithms.}
  \label{fig:silhouette_k}
\end{figure}

\textbf{Effect of $K$:} For unconstrained and agglomerative algorithms, silhouette scores~\citep{rousseeuw1987} peak at low $K$ values ($K=3$ or $K=5$), as shown in Figure~\ref{fig:silhouette_k}, suggesting that Israel's political geography is naturally structured into a small number of macro-regions. SA and Louvain produce lower silhouette scores across all $K$, reflecting the cost of enforcing contiguity and population balance constraints.

\subsection{Best Configurations}

Table~\ref{tab:best} presents selected configurations: the best Agglomerative and K-Means results, the best NMF\_5 Agglomerative result, and the primary Louvain and SA configurations.

\begin{table}[htbp]
  \centering
  \caption{Selected configurations.}
  \label{tab:best}
  \begin{tabular}{clllccc}
    \toprule
    Rank & Repr & Metric & Algorithm & $K$ & Silhouette & Pop CV \\
    \midrule
    1 & BlocShares & Cosine & Agglomerative & 3 & 0.987 & 1.41 \\
    2 & BlocShares & Euclidean & K-Means & 3 & 0.861 & 0.52 \\
    3 & NMF\_5 & Cosine & Agglomerative & 3 & 0.664 & 1.13 \\
    4 & NMF\_5 & Cosine & Louvain & 5 & 0.142 & 0.66 \\
    5 & NMF\_5 & Cosine & SA & 5 & $-0.104^*$ & 0.16$^*$ \\
    \bottomrule
  \end{tabular}

  \smallskip
  {\footnotesize $^*$SA values are 30-seed means (single run: Sil=$-0.109$, CV=0.20); see Section~\ref{sec:limitations} for seed sensitivity analysis.}
\end{table}

\subsection{Primary Result: \texorpdfstring{$K=5$}{K=5} Louvain Partition}

We select $K=5$ with NMF\_5/Cosine/Louvain as the primary result for interpretive analysis based on four criteria (Table~\ref{tab:selection}):

\begin{table}[htbp]
  \centering
  \caption{Selection criteria for primary result. $K=3$ Agglom is BlocShares/Cosine. $^\dagger$Stability is measured at $K=5$ for every configuration; this is BlocShares/Cosine/Agglomerative at $K=5$, the nearest tested configuration.}
  \label{tab:selection}
  \begin{tabular}{lccl}
    \toprule
    Criterion & $K=3$ Agglom & $K=5$ Louvain & Winner \\
    \midrule
    Silhouette (higher = better) & 0.987 & 0.142 & $K=3$ \\
    Population Balance CV (lower = better) & 1.41 & 0.66 & $K=5$ \\
    Temporal Stability ARI (higher = better) & 0.954$^\dagger$ & 0.927 & $K=3$ \\
    Political Granularity ($5 > 3$ cantons) & 3 & 5 & $K=5$ \\
    \bottomrule
  \end{tabular}
\end{table}

\begin{figure}[htbp]
  \centering
  \includegraphics[height=0.85\textheight,keepaspectratio]{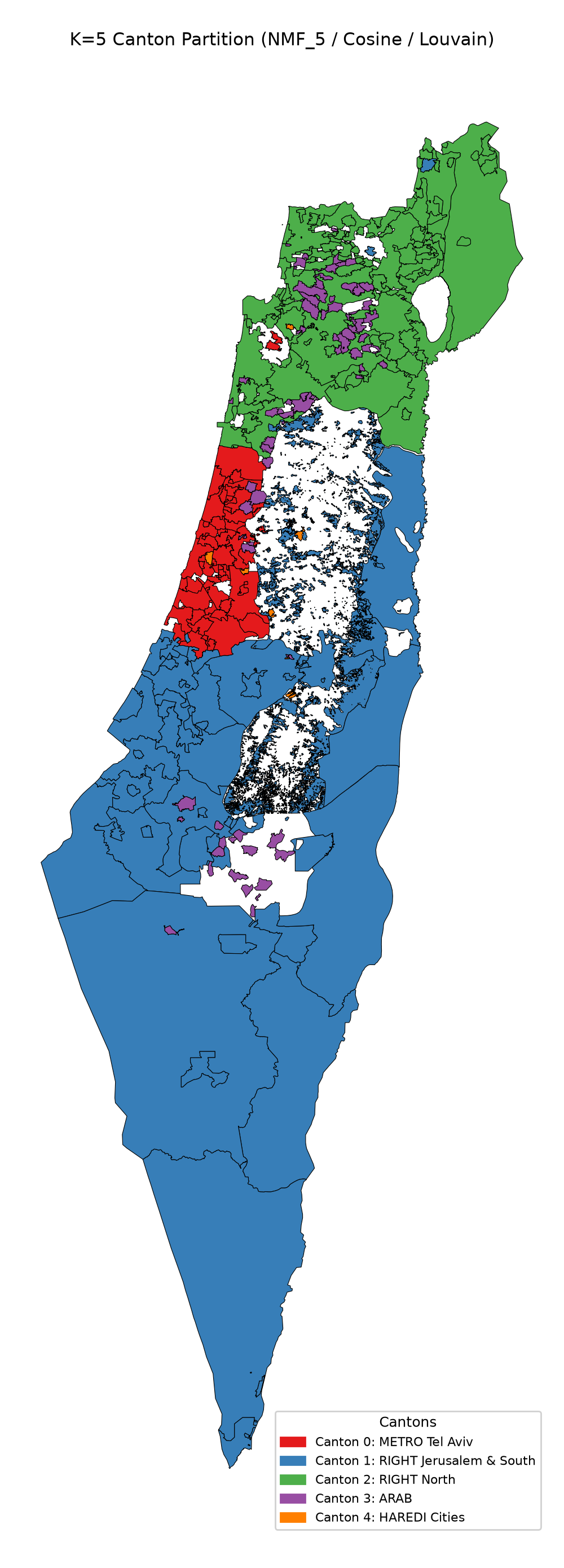}
  \caption{Geographic visualization of the $K=5$ Louvain canton partition (NMF\_5 / Cosine).}
  \label{fig:canton_map}
\end{figure}

\begin{figure}[htbp]
  \centering
  \includegraphics[width=0.8\linewidth]{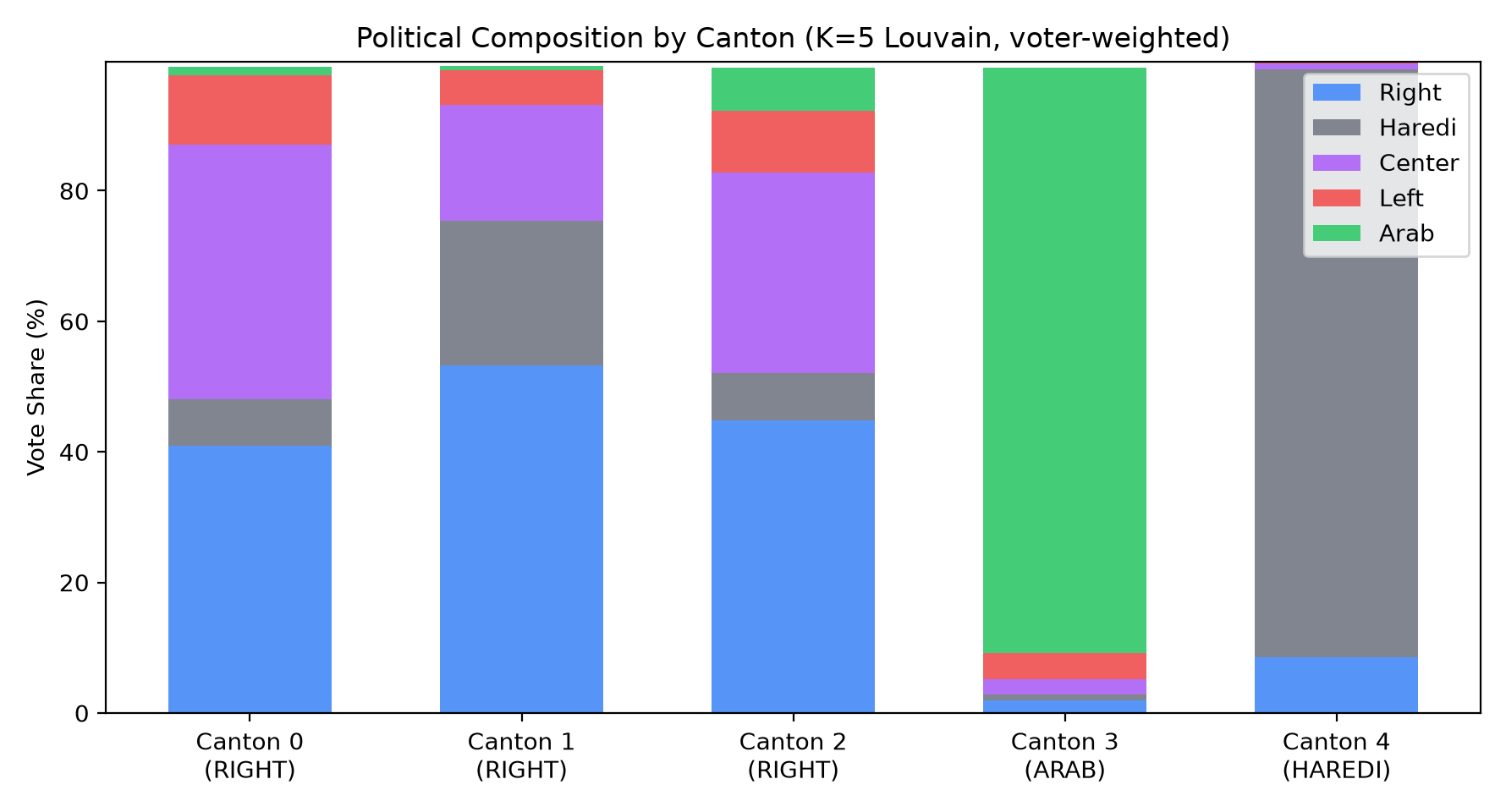}
  \caption{Political bloc composition of each canton in the $K=5$ Louvain partition.}
  \label{fig:composition}
\end{figure}

The criteria split 2--2: $K=3$ Agglomerative has the higher silhouette and stability, $K=5$ Louvain the better population balance and political resolution. We select $K=5$ because the $K=3$ partition achieves its high silhouette by collapsing Israel into three macro-regions, which is well-separated but too coarse to capture the politically meaningful distinctions within the Jewish majority, and its population imbalance is severe (CV = 1.41). The positive silhouette of $K=5$ (0.142) indicates weak but above-chance cluster structure, while providing the granularity needed to distinguish the metropolitan core, two right-leaning regions, the Arab canton and the Haredi cities. $K=6$ was also examined: it splits the Arab canton in two (42 and 18 municipalities) and moves eight towns from RIGHT Jerusalem \& South to the metropolitan canton; its silhouette is slightly higher (0.149) but its population balance (CV 0.90) and cross-election stability (ARI 0.847) are worse. Table~\ref{tab:cantons} presents the canton profiles, Figure~\ref{fig:canton_map} shows the geographic partition, and Figure~\ref{fig:composition} displays the political bloc composition of each canton.

\begin{table}[htbp]
  \centering
  \caption{$K=5$ canton profiles (NMF\_5 / Cosine / Louvain). Bloc values are vote shares (\%) averaged over the five elections and weighted by each municipality's number of voters.}
  \label{tab:cantons}
  \small\setlength{\tabcolsep}{4pt}
  \begin{tabular}{lcccccc}
    \toprule
    Canton & Munis & Right\% & Haredi\% & Center\% & Left\% & Arab\% \\
    \midrule
    0 -- METRO Tel Aviv & 52 & 41.0 & 7.0 & 39.1 & 10.6 & 1.3 \\
    1 -- RIGHT Jerusalem \& South & 61 & 53.2 & 22.2 & 17.7 & 5.3 & 0.7 \\
    2 -- RIGHT North & 75 & 44.8 & 7.3 & 30.7 & 9.4 & 6.6 \\
    3 -- ARAB & 60 & 1.9 & 0.9 & 2.4 & 4.0 & 89.6 \\
    4 -- HAREDI Cities & 7 & 8.6 & 90.0 & 0.9 & 0.3 & 0.0 \\
    \bottomrule
  \end{tabular}
\end{table}

Canton 0 (METRO Tel Aviv) covers Greater Tel Aviv and the Sharon, from Netanya to Yavne and Gezer (plus Daliyat al-Karmel and Isfiya on the Carmel, joined through a bridge edge), with Right and Center almost level (41.0\% and 39.1\%). Canton 1 (RIGHT Jerusalem \& South) runs from Jerusalem through the southern periphery and the Negev and includes the Judea and Samaria authorities (53.2\% Right, 22.2\% Haredi). Canton 2 (RIGHT North) is a mixed northern region including Haifa, the Galilee Jewish towns and nine Galilee towns where the Arab vote is largest (44.8\% Right overall). Canton 3 (ARAB) brings together the Arab municipalities of the Galilee, the Triangle and the Negev (89.6\% Arab vote share). Canton 4 (HAREDI Cities) consists of the seven municipalities where Haredi parties receive more than 70\% of the vote---Bnei Brak, El'ad, Rekhasim, Kiryat Ye'arim, Modi'in Illit, Beitar Illit and Immanuel (90.0\% Haredi). None of the seven borders another; they are joined by the graph's enclave edges, so this canton is a political rather than a geographic unit. It appears once Beitar Illit, Modi'in Illit and Immanuel are included, and the two Arab cantons of the first version merge to keep $K=5$.

\section{Stability Analysis}
\label{sec:stability}

\subsection{Methodology}

To assess whether canton boundaries are structurally stable, we perform temporal stability analysis. For a fixed configuration, we run clustering independently on each of the five elections and measure pairwise agreement using the Adjusted Rand Index (ARI)~\citep{hubert1985} and Normalized Mutual Information (NMI)~\citep{strehl2002}.

\subsection{Results}

\begin{table}[htbp]
  \centering
  \caption{Cross-election stability for seven representative configurations ($K=5$; $^{*}$primary partition).}
  \label{tab:stability}
  \begin{tabular}{lllcccc}
    \toprule
    Repr & Metric & Algorithm & Mean ARI & Std ARI & Mean NMI & Std NMI \\
    \midrule
    BlocShares & Euclidean & Louvain & 0.992 & 0.007 & 0.988 & 0.010 \\
    BlocShares & Cosine & Agglomerative & 0.954 & 0.059 & 0.945 & 0.071 \\
    NMF\_5 & Cosine & Louvain$^{*}$ & 0.927 & 0.048 & 0.930 & 0.044 \\
    BlocShares & Euclidean & SA & 0.587 & 0.103 & 0.636 & 0.083 \\
    NMF\_5 & Euclidean & SA & 0.460 & 0.191 & 0.554 & 0.148 \\
    PCA\_5 & Euclidean & SA & 0.420 & 0.144 & 0.527 & 0.111 \\
    RawParty & Cosine & SA & 0.419 & 0.111 & 0.530 & 0.089 \\
    \bottomrule
  \end{tabular}
\end{table}

\subsection{Interpretation}

\textbf{Deterministic algorithms produce highly stable partitions} (Table~\ref{tab:stability}). The BlocShares Louvain configuration produces near-identical partitions across elections (ARI = 0.992) because it optimizes graph modularity on the fixed adjacency structure with edge weights that change only slightly across elections; this near-perfect stability reflects algorithmic insensitivity to small feature-space variations rather than an independent confirmation of geographic structure. The primary NMF\_5/Cosine Louvain partition is somewhat less stable (ARI = 0.927) but still more stable than any SA configuration. Agglomerative clustering produces near-perfect stability (ARI = 0.954), which is the stronger evidence for structural persistence since it operates directly on feature-space distances that vary across elections.

\textbf{SA stability varies by representation.} BlocShares produces the most stable SA partitions (ARI = 0.587), followed by NMF (0.460), PCA (0.420) and RawParty (0.419); the three party-level representations are close to one another.

\begin{figure}[htbp]
  \centering
  \includegraphics[width=0.8\linewidth]{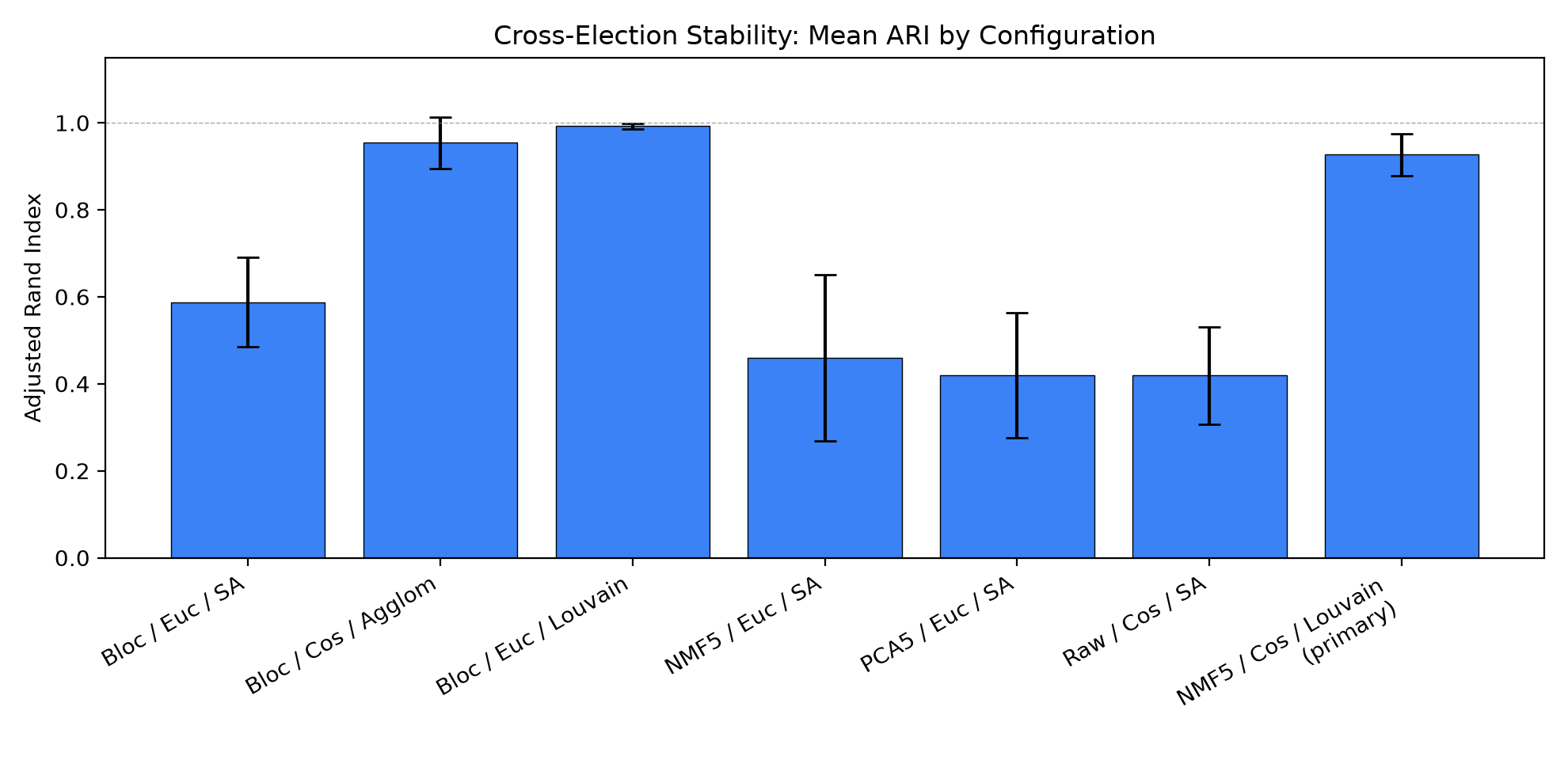}
  \caption{Mean ARI across election pairs for each configuration, with error bars showing standard deviation.}
  \label{fig:stability}
\end{figure}

\textbf{Overall finding:} Over the 2019--2022 period, Israel's political geography remained structurally consistent across elections (Figure~\ref{fig:stability}). The five elections, despite producing different coalition outcomes, exhibit remarkably similar spatial patterns at the municipality level---consistent with the observation that political preferences are strongly spatially autocorrelated~\citep{anselin1995}.

\subsection{Cantons in Each Election}
\label{sec:perelection}

To show the stability of the primary partition directly, the primary configuration was run on each election alone and each election's cantons were matched one-to-one to the primary map's (maximum overlap). A canton sharing less than half of its members with its match (Jaccard index below 0.5) keeps its own number instead of borrowing a label.

\begin{figure}[htbp]
  \centering
  \includegraphics[width=\linewidth]{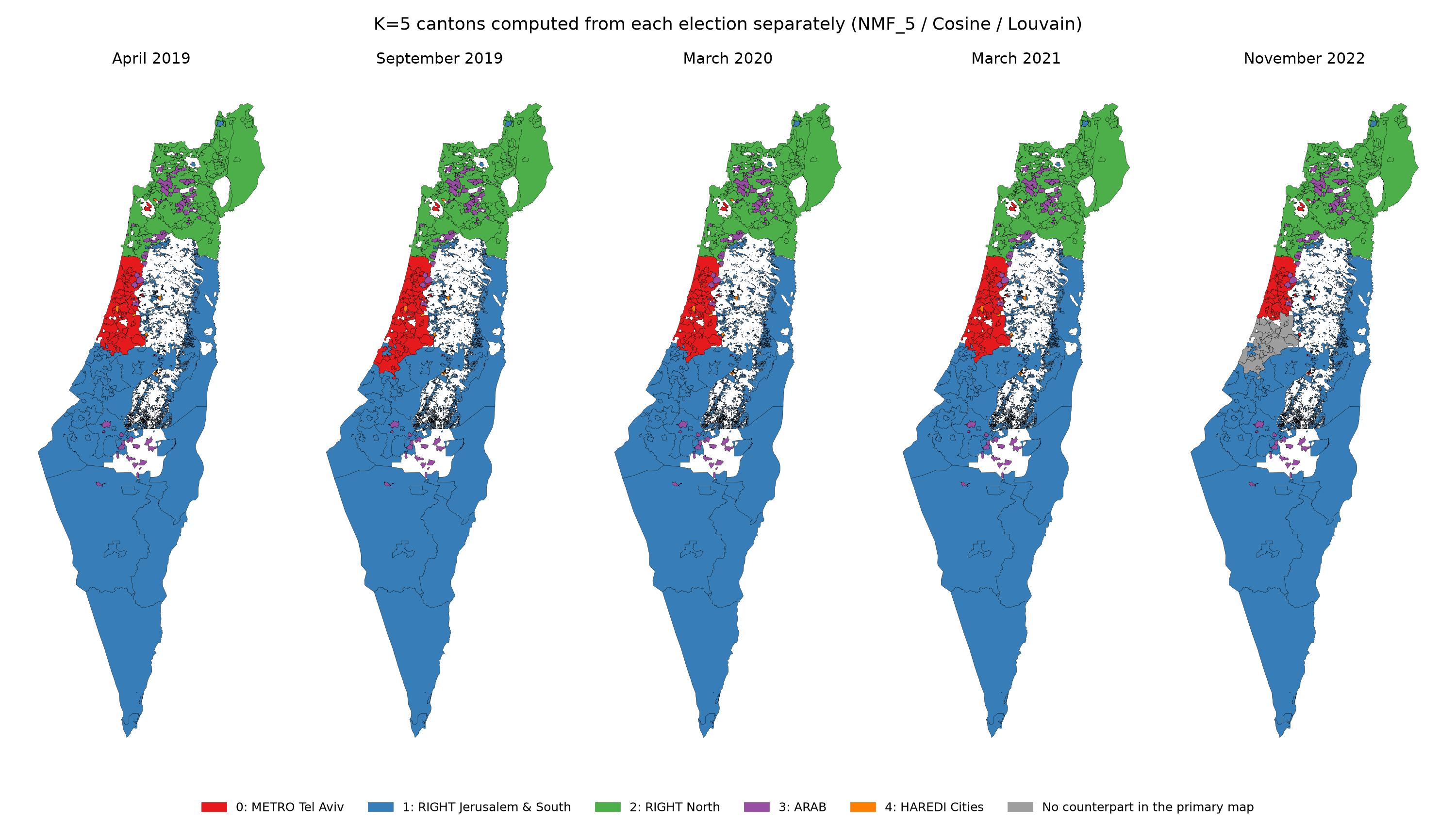}
  \caption{$K=5$ cantons computed from each election separately, coloured by their matching primary canton; grey marks a canton with no counterpart in the primary map (November 2022), in which the Haredi cities join the metropolitan canton.}
  \label{fig:perelection}
\end{figure}

219 of the 255 municipalities are in the same canton in every election; 28 change in one election, 7 in two and one (Nahal Sorek) in four (Figure~\ref{fig:perelection}). April 2019 gives exactly the primary map; in September 2019 eight towns of the southern coastal plain (among them Ashdod, Gedera and Kiryat Malakhi) join the metropolitan canton, and in 2020 and 2021 only two towns and one town move. November 2022 differs most: the seven Haredi cities join the metropolitan canton, and 28 municipalities of the southern metropolitan area and the coastal plain below it (among them Rishon LeZion, Holon, Bat Yam, Rehovot, Ashdod and Modi'in-Maccabim-Re'ut; 20 from the metropolitan canton and 8 from RIGHT Jerusalem \& South) form a canton of their own with no counterpart in the primary map. In 2022 these towns voted 48.8\% Right and 31.1\% Center, against 35.9\% and 42.5\% in the rest of the metropolitan canton.

\subsection{Do the Cantons Differ? A Held-Out Test}
\label{sec:heldout}

Testing whether cantons differ in the vote shares they were built from would be circular: a clustering algorithm separates groups on exactly those features. We therefore rebuild the cantons from the 2019--2021 elections only (features, graph preprocessing and weights all use Knesset 21--24) and test differences on the November 2022 results, which did not shape the map. The training map agrees closely with the primary map (ARI = 0.956), and a unit test checks that scrambling the 2022 results leaves it unchanged.

For each bloc, a Kruskal--Wallis test compares the five cantons, and each pair of cantons is compared by the voter-weighted difference in bloc share with a 95\% bootstrap interval (10,000 resamples of municipalities within each canton) and by a Mann--Whitney test of whether the typical municipality differs, Holm-adjusted across the ten pairs. Each municipality counts once in the Mann--Whitney test, so the two can disagree when a few large towns drive a canton's share.

\begin{table}[htbp]
  \centering
  \caption{Kruskal--Wallis tests across the five cantons, November 2022 bloc shares.}
  \label{tab:heldout}
  \begin{tabular}{lcccc}
    \toprule
    Bloc & $H$ & $p$ & $\epsilon^2$ & Pairs differing (Holm $p<0.05$) \\
    \midrule
    Center & 158.3 & $<10^{-30}$ & 0.62 & 9 of 10 \\
    Arab & 158.0 & $<10^{-30}$ & 0.62 & 9 of 10 \\
    Right & 148.3 & $<10^{-30}$ & 0.58 & 9 of 10 \\
    Haredi & 109.5 & $<10^{-20}$ & 0.43 & 8 of 10 \\
    Left & 88.8 & $<10^{-15}$ & 0.35 & 9 of 10 \\
    \bottomrule
  \end{tabular}
\end{table}

All five blocs differ strongly across cantons in the held-out election (Table~\ref{tab:heldout}; $\epsilon^2$ from 0.35 to 0.62), and 44 of the 50 comparisons (ten canton pairs in each of five blocs) differ significantly. The pairs that do not are informative: the metropolitan canton and RIGHT North do not differ in their typical town's Right, Left or Haredi share, so the line between them is drawn by the Center vote (8.2 points higher in the metropolitan canton) and the Arab vote (6.0 points higher in RIGHT North); and the Haredi share of the metropolitan canton and of RIGHT Jerusalem \& South differs in voters' terms (16.4 points, 95\% interval 5.5 to 24.7, driven by Jerusalem) but not in the typical town ($p = 0.24$).

\textbf{What the test does and does not show.} Municipal bloc shares are very stable between elections: their 2022 values correlate 0.91 (Left) to 0.99 (Arab, Haredi) with the 2019--2021 averages. The held-out test therefore shows that the differences between cantons persist into an election that played no part in drawing them; it does not show that these particular boundaries separate the electorate better than other groupings of the same towns would. In addition, the primary configuration was chosen from a grid computed on all five elections, the bootstrap interval for the seven-town Haredi canton (dominated by Bnei Brak) is likely too narrow, and neighbouring municipalities vote alike, so these tests, which treat municipalities as independent, are somewhat optimistic.

\section{Case Studies}
\label{sec:casestudies}

We examine seven case study areas to assess whether the algorithmic cantons align with known political geography. Dominant blocs and shares are averages over the five elections.

\subsection{Greater Tel Aviv (Gush Dan)}

The whole metropolitan area---the center-leaning core (Tel Aviv, Ramat Gan, Givatayim, Herzliya, Ra'anana, Kfar Saba) and the right-leaning cities around it (Holon, Bat Yam, Petah Tikva, Rishon LeZion, Netanya)---forms Canton 0 (METRO Tel Aviv), where Right and Center are almost level (41.0\% and 39.1\%). Bnei Brak, in the middle of the metropolis, belongs to Canton 4 (HAREDI Cities).

\subsection{Jerusalem Region}

Jerusalem is placed in Canton 1 (RIGHT Jerusalem \& South) with Beit Shemesh, Mevasseret Zion, Ma'ale Adumim, Givat Ze'ev, the regional councils of Judea and Samaria, the southern periphery towns and the major southern cities (53.2\% Right, 22.2\% Haredi). Modi'in joins the metropolitan canton, and Beitar Illit and Modi'in Illit join Canton 4.

\subsection{Judea and Samaria}

Most of the 23 Judea and Samaria authorities join Canton 1, with some of the highest Right-bloc shares in the country (above 85\% in Elkana, Kiryat Arba, Har Hevron, Efrat and Karnei Shomron; Ariel 77.9\%). Oranit and Alfei Menashe, on the Green Line next to the Sharon, join the metropolitan canton, and the Haredi cities Beitar Illit, Modi'in Illit and Immanuel join Canton 4. Because the regional councils' jurisdictions are fragmented and Ariel and Elkana are linked to their nearest authority, several cantons here are held together by links rather than shared borders, as is also the case for the Arab and Haredi cantons inside the Green Line.

\subsection{Arab Towns}

60 of the 69 municipalities where the Arab vote is largest, in the Galilee, the Triangle and the Negev, form Canton 3 (ARAB, 89.6\% Arab vote share); the other nine, Galilee towns with mixed Druze, Christian and Muslim or Bedouin populations such as Maghar, Rama, Abu Snan and Jish, join RIGHT North. The first version of this work split the Arab canton into a Galilee canton and a southern one~\citep{shamir1999}; with the Haredi cities forming their own canton, $K=5$ leaves room for one Arab canton, and it stays whole in each election taken alone (Section~\ref{sec:perelection}).

\subsection{The Haredi Cities}

Canton 4 is the set of municipalities where Haredi parties receive more than 70\% of the vote: Bnei Brak (89.2\%), El'ad (84.8\%), Rekhasim (88.1\%), Kiryat Ye'arim (87.8\%), Modi'in Illit (97.6\%), Beitar Illit (90.7\%) and Immanuel (76.8\%). None of the seven borders another: the canton is held together by the graph's enclave edges, which connect municipalities dominated by the same bloc. It is a political archipelago rather than a geographic region, to be read as ``the Haredi cities vote as a bloc of their own'' rather than as a contiguous territory.

\subsection{Haifa and the Krayot}

Haifa and all the Krayot are assigned to Canton 2 (RIGHT North, 44.8\% Right), a large northern canton encompassing the Jewish towns of the Galilee and upper Galilee alongside the Haifa metropolitan area.

\subsection{Southern Periphery}

All southern periphery towns---Beer Sheva, Dimona, Netivot, Ofakim, Arad, Eilat---are grouped in Canton 1 (RIGHT Jerusalem \& South), forming a contiguous arc from Jerusalem through the Negev to Eilat. This aligns with the well-documented ``periphery vote'' phenomenon~\citep{arian2008, shamir1999}.

\subsection{Comparison to Administrative Districts}

Assigning each municipality the most common district among its localities, the ARI~\citep{hubert1985} between our $K=5$ political cantons and Israel's official administrative districts is 0.354 over all 255 municipalities (the first version reported 0.435 for an earlier map, computed only over cities and local councils). The Arab canton spans five districts, the Judea and Samaria district is split three ways, and Jerusalem groups with the southern periphery: political cantons follow ideological lines rather than administrative boundaries.

\section{Limitations and Future Work}
\label{sec:limitations}

\subsection{Limitations}

\textbf{Population Imbalance:} The $K=5$ partition has a municipality count range of 7--75 (Pop CV = 0.66); the Haredi canton is far smaller than the others. Louvain optimizes modularity rather than balance.

\textbf{SA Stochasticity:} SA results in the grid search use 5,000 iterations per configuration with a fixed seed, which makes each run reproducible. A 30-seed sensitivity analysis of the primary SA configuration (NMF\_5/Cosine/$K=5$) using 50,000 iterations showed silhouette scores ranging from $-0.215$ to $-0.055$ (mean: $-0.104$, std: 0.026) and Pop CV from 0.01 to 0.21 (mean: 0.16, std: 0.08). This variance is smaller than the gap between SA and Louvain, confirming that algorithmic choice dominates over seed variance.

\textbf{Ecological Fallacy:} Our analysis operates at the municipality level, aggregating individual votes~\citep{robinson1950}. Within-municipality variation is not captured.

\textbf{Static Analysis:} We analyze election results as fixed snapshots without modeling demographic shifts or political realignment over time.

\textbf{Data Coverage:} The 255 municipalities with votes in all five elections are included, among them the 23 Israeli local authorities in Judea and Samaria; small localities within regional councils are aggregated at the council level. The Judea and Samaria boundaries combine CBS and Civil Administration sources, neither of which states terms of use.

\textbf{Contiguity Approximation:} Virtual and enclave edges mean some cantons are not strictly contiguous: Kiryat Shmona, in the far north, joins Canton 1 through enclave edges to Sderot and Dimona; Beit Jann, which shares no border with any municipality in the data, joins Canton 1 through virtual edges; Daliyat al-Karmel and Isfiya join the metropolitan canton through a bridge edge; and the Haredi canton's seven municipalities share no borders. Only RIGHT North is a single piece on the map.

\textbf{Selection Bias:} With 264 configurations evaluated, reported ``best'' silhouette scores may be inflated by selection effects. Relative rankings and qualitative patterns are more robust than absolute performance differences.

\subsection{Future Work}

\begin{itemize}
  \item \textbf{Finer geographic resolution:} Extend analysis to statistical area or neighborhood level within large cities.
  \item \textbf{Coalition simulation:} Use canton-level profiles to simulate coalition formation under hypothetical regional representation.
  \item \textbf{Cross-country comparison:} Apply the methodology to other countries with spatial political polarization (e.g., United States, Belgium).
  \item \textbf{Temporal dynamics:} Model how canton boundaries shift over time as demographics change.
  \item \textbf{Parameter sensitivity:} Systematic sensitivity analysis of SA cost function weights and graph preprocessing thresholds.
\end{itemize}

\section{Ethical Considerations}
\label{sec:ethics}

This work is purely descriptive and exploratory. The ``canton'' partitions produced by our algorithms are analytical constructs designed to reveal spatial patterns in voting data; they do not constitute a political proposal or recommendation for territorial division. All data used are publicly available, aggregated at the municipality level, and contain no individual-level information. Readers should be aware of the ecological fallacy~\citep{robinson1950}: municipality-level vote shares do not reflect the preferences of any individual voter. The analysis includes the Israeli local authorities in Judea and Samaria because their residents vote in Knesset elections; Palestinian areas, which do not, are outside it. Their inclusion reflects where Knesset votes are cast and implies no position on the status of the territories.

\section{Conclusion}
\label{sec:conclusion}

This paper demonstrates that constrained spatial clustering can partition Israeli municipalities into politically coherent cantons of neighbouring municipalities that align with known political-demographic divisions; only one of the five is a single piece on the map, because the analysis graph also links politically similar towns that share no borders. Our systematic experimental framework---evaluating 264 configurations---provides comprehensive evidence that:

\begin{enumerate}
  \item \textbf{Israel's political geography is structurally robust.} BlocShares achieves the highest silhouette scores for unconstrained algorithms, while NMF\_5 with Louvain produces the most interpretable partition with positive silhouette (0.142) and stable boundaries across elections (ARI = 0.927).

  \item \textbf{The Arab-Jewish divide is the dominant political-geographic cleavage, with the Haredi cities a second bloc of their own.} In the primary partition 60 of the 69 municipalities where the Arab vote is largest form one canton; with the Judea and Samaria authorities included, the seven predominantly Haredi cities also form their own canton, although they share no borders.

  \item \textbf{Political cantons differ from administrative districts.} With ARI = 0.354 between our $K=5$ cantons and administrative districts, political geography follows ideological lines rather than administrative boundaries.

  \item \textbf{The differences between cantons persist into a new election.} Built from the 2019--2021 elections only, the cantons still differ significantly in all five blocs' vote shares in the held-out 2022 election (44 of 50 comparisons of canton pairs, $\epsilon^2$ 0.35--0.62). Because municipal voting is so stable, this shows persistence rather than validating these exact boundaries.

  \item \textbf{Multi-objective optimization is essential.} Louvain produces the most interpretable partitions with the best balance of silhouette, stability, and political coherence among community detection methods, while SA achieves the best population balance through its explicit multi-objective cost function.

  \item \textbf{Distance metric choice matters.} Cosine distance with BlocShares/Agglomerative at $K=3$ achieves the highest silhouette score (0.987), ahead of Euclidean (0.910), while K-Means partitions that score well under Euclidean distance score poorly under Cosine and Jensen-Shannon.
\end{enumerate}

This work bridges computer science, geographic information systems, and political analysis, providing both a methodological contribution to constrained spatial clustering and an analytical tool for exploring Israel's political geography.

\section*{Acknowledgments}

This work was conducted as part of the first author's M.Sc.\ project at the Department of Computer Science, Bar-Ilan University. We thank the Israel Central Bureau of Statistics for making election results and geographic boundary data publicly available.

\bibliographystyle{unsrtnat}
\bibliography{references}

@article{stephanopoulos2015,
  author    = {Stephanopoulos, Nicholas O. and McGhee, Eric M.},
  title     = {Partisan Gerrymandering and the Efficiency Gap},
  journal   = {University of Chicago Law Review},
  volume    = {82},
  number    = {2},
  pages     = {831--900},
  year      = {2015}
}

@misc{brieden2017,
  author    = {Brieden, Andreas and Gritzmann, Peter and Klemm, Fabian},
  title     = {Constrained clustering via diagrams: A unified theory and its application to electoral district design},
  howpublished = {arXiv preprint arXiv:1703.02867},
  year      = {2017}
}

@article{deford2021,
  author    = {DeFord, Daryl and Duchin, Moon and Solomon, Justin},
  title     = {Recombination: A family of {M}arkov chains for redistricting},
  journal   = {Harvard Data Science Review},
  volume    = {3},
  number    = {1},
  year      = {2021}
}

@misc{duchin2018,
  author    = {Duchin, Moon and Tenner, Bridget Eileen},
  title     = {Discrete geometry for electoral geography},
  howpublished = {arXiv preprint arXiv:1808.05860},
  year      = {2018}
}

@book{hastie2009,
  author    = {Hastie, Trevor and Tibshirani, Robert and Friedman, Jerome},
  title     = {The Elements of Statistical Learning: Data Mining, Inference, and Prediction},
  publisher = {Springer},
  edition   = {2nd},
  year      = {2009}
}

@article{murtagh2014,
  author    = {Murtagh, Fionn and Legendre, Pierre},
  title     = {Ward's hierarchical agglomerative clustering method: Which algorithms implement {W}ard's criterion?},
  journal   = {Journal of Classification},
  volume    = {31},
  number    = {3},
  pages     = {274--295},
  year      = {2014}
}

@article{luxburg2007,
  author    = {von Luxburg, Ulrike},
  title     = {A tutorial on spectral clustering},
  journal   = {Statistics and Computing},
  volume    = {17},
  number    = {4},
  pages     = {395--416},
  year      = {2007}
}

@article{blondel2008,
  author    = {Blondel, Vincent D. and Guillaume, Jean-Loup and Lambiotte, Renaud and Lefebvre, Etienne},
  title     = {Fast unfolding of communities in large networks},
  journal   = {Journal of Statistical Mechanics: Theory and Experiment},
  volume    = {2008},
  number    = {10},
  pages     = {P10008},
  year      = {2008},
  doi       = {10.1088/1742-5468/2008/10/P10008}
}

@article{kirkpatrick1983,
  author    = {Kirkpatrick, Scott and Gelatt, C. Daniel and Vecchi, Mario P.},
  title     = {Optimization by simulated annealing},
  journal   = {Science},
  volume    = {220},
  number    = {4598},
  pages     = {671--680},
  year      = {1983},
  doi       = {10.1126/science.220.4598.671}
}

@article{arian2008,
  author    = {Arian, Asher and Shamir, Michal},
  title     = {A decade later, the world had changed, the cleavage structure remained: {I}srael 1996--2006},
  journal   = {Party Politics},
  volume    = {14},
  number    = {6},
  pages     = {685--705},
  year      = {2008}
}

@book{jolliffe2002,
  author    = {Jolliffe, Ian T.},
  title     = {Principal Component Analysis},
  publisher = {Springer},
  edition   = {2nd},
  year      = {2002}
}

@article{lee1999,
  author    = {Lee, Daniel D. and Seung, H. Sebastian},
  title     = {Learning the parts of objects by non-negative matrix factorization},
  journal   = {Nature},
  volume    = {401},
  number    = {6755},
  pages     = {788--791},
  year      = {1999},
  doi       = {10.1038/44565}
}

@article{rousseeuw1987,
  author    = {Rousseeuw, Peter J.},
  title     = {Silhouettes: A graphical aid to the interpretation and validation of cluster analysis},
  journal   = {Journal of Computational and Applied Mathematics},
  volume    = {20},
  pages     = {53--65},
  year      = {1987},
  doi       = {10.1016/0377-0427(87)90125-7}
}

@article{hubert1985,
  author    = {Hubert, Lawrence and Arabie, Phipps},
  title     = {Comparing partitions},
  journal   = {Journal of Classification},
  volume    = {2},
  number    = {1},
  pages     = {193--218},
  year      = {1985}
}

@article{strehl2002,
  author    = {Strehl, Alexander and Ghosh, Joydeep},
  title     = {Cluster ensembles -- a knowledge reuse framework for combining multiple partitions},
  journal   = {Journal of Machine Learning Research},
  volume    = {3},
  pages     = {583--617},
  year      = {2002}
}

@article{newman2004,
  author    = {Newman, Mark E. J. and Girvan, Michelle},
  title     = {Finding and evaluating community structure in networks},
  journal   = {Physical Review E},
  volume    = {69},
  number    = {2},
  pages     = {026113},
  year      = {2004},
  doi       = {10.1103/PhysRevE.69.026113}
}

@article{endres2003,
  author    = {Endres, Dominik M. and Schindelin, Johannes E.},
  title     = {A new metric for probability distributions},
  journal   = {IEEE Transactions on Information Theory},
  volume    = {49},
  number    = {7},
  pages     = {1858--1860},
  year      = {2003}
}

@article{lin1991,
  author    = {Lin, Jianhua},
  title     = {Divergence measures based on the {S}hannon entropy},
  journal   = {IEEE Transactions on Information Theory},
  volume    = {37},
  number    = {1},
  pages     = {145--151},
  year      = {1991}
}

@article{robinson1950,
  author    = {Robinson, William S.},
  title     = {Ecological correlations and the behavior of individuals},
  journal   = {American Sociological Review},
  volume    = {15},
  number    = {3},
  pages     = {351--357},
  year      = {1950}
}

@article{pedregosa2011,
  author    = {Pedregosa, Fabian and Varoquaux, Ga{\"e}l and Gramfort, Alexandre and Michel, Vincent and Thirion, Bertrand and Grisel, Olivier and Blondel, Mathieu and Prettenhofer, Peter and Weiss, Ron and Dubourg, Vincent and others},
  title     = {Scikit-learn: Machine learning in {P}ython},
  journal   = {Journal of Machine Learning Research},
  volume    = {12},
  pages     = {2825--2830},
  year      = {2011}
}

@inproceedings{hagberg2008,
  author    = {Hagberg, Aric A. and Schult, Daniel A. and Swart, Pieter J.},
  title     = {Exploring network structure, dynamics, and function using {N}etwork{X}},
  booktitle = {Proceedings of the 7th Python in Science Conference (SciPy)},
  pages     = {11--15},
  year      = {2008}
}

@article{virtanen2020,
  author    = {Virtanen, Pauli and Gommers, Ralf and Oliphant, Travis E. and Haberland, Matt and Reddy, Tyler and Cournapeau, David and Burovski, Evgeni and Peterson, Pearu and Weckesser, Warren and Bright, Jonathan and others},
  title     = {{S}ci{P}y 1.0: Fundamental algorithms for scientific computing in {P}ython},
  journal   = {Nature Methods},
  volume    = {17},
  pages     = {261--272},
  year      = {2020},
  doi       = {10.1038/s41592-019-0686-2}
}

@article{assuncao2006,
  author    = {Assun{\c{c}}{\~a}o, Renato M. and Neves, Marcos Corr{\^e}a and C{\^a}mara, Gilberto and da Costa Freitas, Corina},
  title     = {Efficient regionalization techniques for socio-economic geographical units using minimum spanning trees},
  journal   = {International Journal of Geographical Information Science},
  volume    = {20},
  number    = {7},
  pages     = {797--811},
  year      = {2006}
}

@article{diskin2005,
  author    = {Diskin, Abraham and Diskin, Hanna},
  title     = {The politics of electoral reform in {I}srael},
  journal   = {International Political Science Review},
  volume    = {26},
  number    = {1},
  pages     = {33--54},
  year      = {2005}
}

@article{shamir1999,
  author    = {Shamir, Michal and Arian, Asher},
  title     = {Collective identity and electoral competition in {I}srael},
  journal   = {American Political Science Review},
  volume    = {93},
  number    = {2},
  pages     = {265--277},
  year      = {1999}
}

@misc{cbs2023elections,
  author    = {{Israel Central Bureau of Statistics}},
  title     = {Elections, the {K}nesset, and Government -- Statistical Abstract},
  year      = {2023},
  url       = {https://www.cbs.gov.il/en/subjects/Pages/Elections-and-the-Knesset.aspx},
  note      = {Accessed: 2024}
}

@misc{cbs2023geo,
  author    = {{Israel Central Bureau of Statistics}},
  title     = {Geographic Layers -- Municipal Boundaries {GIS} Data},
  year      = {2023},
  url       = {https://www.cbs.gov.il/he/Pages/geo-layers.aspx},
  note      = {Accessed: 2024}
}

@book{bellman1961,
  author    = {Bellman, Richard E.},
  title     = {Adaptive Control Processes: A Guided Tour},
  publisher = {Princeton University Press},
  year      = {1961}
}

@article{duque2012,
  author    = {Duque, Juan C. and Anselin, Luc and Rey, Sergio J.},
  title     = {The {MAX-P-REGIONS} problem},
  journal   = {Journal of Regional Science},
  volume    = {52},
  number    = {3},
  pages     = {397--419},
  year      = {2012}
}

@article{anselin1995,
  author    = {Anselin, Luc},
  title     = {Local indicators of spatial association -- {LISA}},
  journal   = {Geographical Analysis},
  volume    = {27},
  number    = {2},
  pages     = {93--115},
  year      = {1995}
}

@misc{strenger2014,
  author       = {Strenger, Carlo and Yadid, Judd},
  title        = {Lama rak shtei medinot? Gam {Israel} yekhola lehitchalek le'otonomiot [Why only two states? {Israel}, too, can be divided into autonomies]},
  howpublished = {Haaretz, The Edge},
  year         = {2014},
  month        = oct,
  note         = {8 October 2014. English edition: J. Yadid and C. Strenger, ``How cantonization can save Israel'', Haaretz, 7 October 2014}
}

@misc{civiladmin2026,
  author       = {{Civil Administration of Judea and Samaria}},
  title        = {Regional council areas and settlement jurisdiction orders ({GIS} layers, freedom of information)},
  year         = {2026},
  note         = {Retrieved 11 October 2026}
}

\end{document}